\documentclass[12pt]{article}

\usepackage{graphicx}
\usepackage{dcolumn}
\usepackage{bm}

\begin{document}


\title{Decoherence induced continuous pointer states\thanks{Dedicated to Rudolf Haag  on his $80^{\rm th}$ birthday}} 

\author{Ph. Blanchard\\Physics Faculty and BiBoS, University of Bielefeld\\33615 Bielefeld \and
R. Olkiewicz\\Institute of Theoretical Physics, University of Wroc{\l}aw\\50204 Wroc{\l}aw}
\date{July 2002}

\maketitle

\begin{abstract}
We investigate the reduced dynamics in the Markovian approximation of an infinite quantum
spin system linearly coupled to a phonon field at positive temperature. The achieved
diagonalization leads to a selection of the continuous family of pointer states
corresponding to a configuration space of the one-dimensional Ising model. Such a family
provides a mathematical description of an apparatus with continuous readings.
\end{abstract}



Decoherence is a process of continuous measurement-like interactions between a quantum
system and its environment which results in limiting the validity of the superposition
principle in the Hilbert space of the system \cite{Zurek2}. It accepts the wave function
description of the combined state of the system and its environment but contends that it
is practically impossible to distinguish it from the corresponding statistical mixture.
In other words, the environment destroys the vast majority of the superpositions quickly,
and, in the case of macroscopic objects, almost instantaneously. It was shown that
decoherence is a universal short time phenomenon independent of the character of the
system and the reservoir \cite{Braun}.

In recent years decoherence has been widely discussed and accepted as a mechanism
responsible for the emergence of classicality in quantum open systems
\cite{Joos,Giul,Blan}. A particular aspect of decoherence is the selection of the
preferred basis of pointer states \cite{Zurek1} which occurs when the reduced density
matrix of the system becomes approximately diagonal in time much shorter than the
relaxation time. In practical situations this results in disappearing of non-diagonal
elements in the reduced density matrix. Hence, by definition, pointer states do not
evolve at all, while all other states deteriorate in time to classical probability
distributions over the one dimensional projections corresponding to these states.
However, it should be pointed out that the algebra generated by these projections is
always of a discrete type, and, as was shown in \cite{Olk}, the discreteness is
unavoidable as long as we consider quantum systems with a finite number of degrees of
freedom. Such a situation is clearly unsatisfactory since there are quantum measurements
with continuous outcomes. In this Letter we demonstrate by a completely solvable model
that "openness" of a macroscopic measuring device, regarded as a quantum system in the
thermodynamic limit, yields continuous pointer states. By continuous pointer states we
understand an uncountable family of commuting and dynamically invariant projections which
contains no minimal projections, and such that any observable of the apparatus evolves
towards the Abelian algebra generated by these projections.

The model is the following (we shall work in the Heisenberg picture). The apparatus is a
semi-infinite linear array of spin-$\frac{1}{2}$ particles, fixed at positions
$n=1,\,2,...\,$. The algebra $\cal M$ of its (bounded) observables is given by the
$\sigma$-weak closure of $\pi(\otimes_1^{\infty}M_{2\times 2})$, where $\pi$ is a
(faithful) GNS representation with respect to a tracial state tr on the Glimm algebra
$\otimes_1^{\infty}M_{2\times 2}$, and $M_{2\times 2}$ is the algebra generated by Pauli
matrices. Since there is no free evolution of the apparatus, $H_A=0$. The reservoir is
chosen to consists of noninteracting phonons of an infinitely extended one dimensional
harmonic crystal at the inverse temperature $\beta =\frac{1}{kT}$. The Hilbert space
$\cal H$ representing pure states of a single phonon is (in the momentum representation)
${\cal H}=L^2({\bf R},\,dk)$. A phonon energy operator is given by the dispersion
relation $\omega(k)=|k|$ ($\hbar =1,\;c=1)$. It follows that the Hilbert space of the
reservoir is ${\cal F}\otimes{\cal F}$, where $\cal F$ is the symmetric Fock space over
$\cal H$. A phonon field $\phi(f)=\frac{1}{\sqrt{2}}(a^*(f)+a(f))$, where $a^*(f)$ and
$a(f)$ are given by the Araki-Woods representation \cite{Araki}:
\begin{equation}
a^*(f)\:=\:a_F^*((1+\rho)^{1/2}f)\otimes I\:+\:I\otimes a_F(\rho^{1/2}\bar{f}),
\end{equation}
\begin{equation}
a(f)\:=\:a_F((1+\rho)^{1/2}f)\otimes I\:+\:I\otimes a_F^*(\rho^{1/2}\bar{f}).
\end{equation}
Here $a_F^*(a_F)$ denotes respectively creation (annihilation) operators in the Fock
space, and $\rho$ is the thermal equilibrium distribution related to the phonons energy
according to the Planck law
\begin{equation}
\rho(k)\:=\:\frac{1}{e^{\beta\omega(k)}-1}.
\end{equation}
Since the phonons are noninteracting, their dynamics is completely determined by the
energy operator
\begin{equation}
H_E\:=\:H_0\otimes I\:-\:I\otimes H_0,
\end{equation}
where $H_0=d\Gamma(\omega)=\int\omega(k)a_F^*(k)a_F(k)dk$ describes dynamics of the
reservoir at zero temperature. The reference state of the reservoir is taken to be a
gauge-invariant quasi-free thermal state given by
\begin{equation}
\omega_E(a^*(f)a(g))\:=\:\int\rho(k)\bar{g}(k)f(k)dk.
\end{equation}
Clearly, $\omega_E$ is invariant with respect to the free dynamics of the environment.

The Hamiltonian $H$ of the joint system consists of the reservoir term $H_E$ and an
interacting part $H_I$. We assume that the coupling is linear (as in the spin-boson
model), i.e. $H_I=\lambda Q\otimes\phi(g)$, where
\begin{equation}
Q\:=\:\pi\left(\sum\limits_{n=1}^{\infty}\frac{1}{2^n}\sigma_n^3\right),
\end{equation}
$\sigma_n^3$ is the third Pauli matrix in the n-th site, and $\lambda>0$ is a coupling
constant. In real interactions one should also include bilinear terms in the coupling.
However, even this simplified model turns out to yield an efficient loss of coherence of
the vast majority of the apparatus observables. The factor $\frac{1}{2^n}$ in Eq. (6)
reflects the property that interaction between spin particles and the reservoir decreases
as $n\to\infty$. Its form was chosen to simplify further calculations. The test function
$g(k)=|k|^{1/2}\chi(k)$, where $\chi(k)$ is an even and real valued  function
such that:\\
(i) $\chi$ is differentiable with bounded derivative,\\
(ii) for large $|k|$, $|\chi(k)|\leq\frac{C}{k^{2+\epsilon}}$, $C>0$, $\epsilon>0$,\\
and $\chi(0)=1$. The behavior of the test function $g$ at the origin and the asymptotic
bound (ii) are taken to ensure that $H$ is essentially self-adjoint. Properties (i) and
(ii) will also secure that the thermal correlation function of the field operator is
integrable (see below). The cutoff function $\chi(k)$ may be of a Gaussian, exponential
or algebraic type. Since
\begin{equation}
H_I\:=\:\frac{\lambda}{\sqrt{2}}Q\int dk\,g(k)(a_k^*\:+\:a_k),
\end{equation}
the spectral density of the environmental coupling
\begin{equation}
J(\omega)\:\sim\:\int dk\,g(k)^2\delta(\omega\:-\:\omega(k))
\end{equation}
is linear for small values of $\omega$. Hence environmental dissipation modelled by by
Eq. (7) corresponds to the so-called ohmic case \cite{Legg}.

The reduced dynamics of an apparatus observable $X$ is given by
\begin{equation}
T_t(X)\:=\:\Pi^{\omega_E}(e^{itH}(X\otimes I)e^{-itH}),
\end{equation}
where $\Pi^{\omega_E}$ is a conditional expectation (the dual operation to the partial
trace) with respect to the reference state $\omega_E$ of the reservoir. We derive an
explicit formula for $T_t$ in the Markovian approximation which proved to be successful
also in other models \cite{Unruh,Twam}. Because the thermal correlation function
\begin{eqnarray}
<\phi_t(g)\phi(g)>\:=\:\omega_E(e^{itH_E}\phi(g)e^{-itH_E}\phi(g))\nonumber\\
=\:\omega_E(\phi(e^{it\omega}g)\phi(g))
\end{eqnarray}
is integrable, we use the so-called singular coupling limit \cite{Ali,Pal} which states
that $T_t=e^{tL}$ is a quantum Markov semigroup with the generator $L$ given by a master
equation in the standard (Gorini-Kossakowski-Sudershan-Lindblad) form
\begin{equation}
L(X)\:=\:ib[X,\,Q^2]\:+\:\lambda a(QXQ\:-\:\{Q^2,\,X\}).
\end{equation}
Parameters $a>0$ and $b\in{\bf R}$ are determined by
\begin{equation}
\int\limits_0^{\infty}<\phi_t(g)\phi(g)>dt\:=\:\frac{a}{2}\:+\:ib.
\end{equation}
By direct calculations
\begin{eqnarray}
<\phi_t(g)\phi(g)>\:=\:\sqrt{2\pi}F(f_1)(t)\:+\:\frac{\sqrt{2\pi}}{2}F(f_2)(t)\nonumber\\
+\:\frac{\sqrt{2\pi}}{2}F(f_3)(t),
\end{eqnarray}
where
\begin{equation}
f_1(k)\:=\:\frac{|k|\chi^2(k)}{e^{\beta|k|}-1},
\end{equation}
\begin{equation}
f_2(k)\:=\:|k|\chi^2(k),\quad f_3(k)\:=\:k\chi^2(k),
\end{equation}
and $F$ stands for the Fourier transform. Hence, by the inverse Fourier formula,
\begin{equation}
a\:=\:2\pi f_1(0)\:+\:\pi f_2(0)\:=\:\frac{2\pi}{\beta},
\end{equation}
and
\begin{equation}
b\:=\:\frac{\sqrt{2\pi}}{2}\int\limits_0^{\infty}\left(\frac{d}{dt}F(\chi^2)(t)\right)
dt\:=\:-\int\limits_0^{\infty}\chi^2(k)dk.
\end{equation}
The master equation (11) consists of two terms. The first one is a Hamiltonian term
$H_A'=bA^2$, and the second is a dissipative operator
\begin{equation}
L_D(X)\:=\:\frac{2\pi\lambda}{\beta}(AXA\:-\:\frac{1}{2}\{A^2,\,X\}).
\end{equation}
Because these two parts commute so, by the Trotter product formula,
\begin{equation}
T_t(X)\:=\:e^{itH_A'}(e^{tL_D}X)e^{-itH_A'}.
\end{equation}
We now describe effects of dissipation. Because $\cal M$ is a limit of local algebras
$M_{2^n\times 2^n}=\otimes_1^{\infty}M_{2\times 2}$, and $e^{tL_D}$ preserves each local
algebra so we may assume that $X=(x_{ij})\in M_{2^n\times 2^n}$. Then
\begin{equation}
L_D(X)_{ij}\:=\:-\frac{\pi\lambda}{\beta}\cdot\frac{(j-i)^2}{4^{n-1}}x_{ij},
\end{equation}
$i,j\in\{1,...,2^n\}$, and so
\begin{equation}
e^{tL_D}(X)_{ij}\:=\:x_{ij}\exp\left(-\gamma t\frac{(j-i)^2}{4^{n-1}}\right),
\end{equation}
where $\gamma =\pi k\lambda T$. It follows from Eq. (20) that the loss of coherence is
faster for coefficients which are more distant to the diagonal, and it increases with
reservoir temperature similarly as in the model of a harmonic oscillator linearly coupled
to an infinite bath of harmonic oscillators \cite{Twam}, and for a spinless quantum
particle minimally coupled to the radiation field \cite{Durr}. In the thermodynamic limit
$n\to\infty$ dissipation leads to an approximate diagonalization of apparatus observables
in any finite time interval. However, the $t\to\infty$ limit leads to the following
result. Suppose $\cal A$ is a von Neumann algebra generated by $\pi(\sigma_n^3)$,
$n\in{\bf N}$. Then $\cal A$ is a maximal commutative subalgebra in $\cal M$, and let
$P:{\cal M}\to {\cal A}$ be a von Neumann projection onto it. Since
$[H_A',\,\sigma_n^3]=0$, it follows
from equations (18) and (20) that all observables from $\cal A$ are $T_t$-invariant.\\
THEOREM: {\it For any statistical state (density matrix) $\Lambda$ of the spin system and
any spin observable} $X$
\begin{equation}
\lim\limits_{t\to\infty}<T_t(X)>_{\Lambda}\:=\:<P(X)>_{\Lambda},
\end{equation}
{\it where $<X>_{\Lambda}=\,{\rm tr}(\Lambda X)$ is the expectation value of $X$ in
state} $\Lambda$.\\
By Eq. (21) all expectation values of $\pi(\sigma^k_n)$, where $k=1,2$, and $n\in{\bf
N}$, tend to zero, and so the $t\to\infty$ limit yields a complete diagonalization of any
spin observable.

Finally, we describe the algebra $\cal A$. Since $\cal A$ is commutative, it is an
algebra of functions on some configuration space $\Omega$. In the sequel we identify an
operator $X\in{\cal A}$ with the corresponding function $X(\eta)$, $\eta\in\Omega$. Let
$P_n^+$ and $P_n^-$ be spectral projections of $\sigma_n^3$, i.e.
$\sigma_n^3=P_n^+\,-\,P_n^-$. An infinite product $P_1^{\sharp}P_2^{\sharp}\cdots$, where
$\sharp$ stands for + or -, defines a state on the subalgebra of continuous functions in
$\cal A$, and so corresponds to a point in the configuration space. Thus
$\Omega=\,\{(i_1,\,i_2,...):\;\;i_n=\pm\}$ or, in other words, each point of $\Omega$
describes a configuration of up and down spins located at $n=1,2,...\,$. If $\mu_0$ is a
probability measure on $\{-1,\,1\}$ which assigns value one half to both $\uparrow$ and
$\downarrow$ spin positions, and if $\mu =\otimes_1^{\infty}\mu_0$ is the corresponding
product probability measure on $\Omega$, then for any $X\in{\cal A}$
\begin{equation}
{\rm tr}(X)\:=\:\int X(\eta)d\mu(\eta),
\end{equation}
and so the induced pointer states form an uncountable family. More precisely, for any
$s\in[0,\,1]$ there exists a projection $e\in{\cal A}$ such that tr$(e)=s$. Thus, since
normalization to the unit interval is not essential, the decoherence induced pointer
states of the presented model indeed correspond to a pointer with continuous readings.

It is worth noting that continuous families of projections have been selected by a
decoherence process also in other models. For example, using the so-called predictability
sieve coherent states of a harmonic oscillator coupled to a heat bath (quantum Brownian
motion) were shown to be the most stable ones \cite{Zurek3}. In a different framework it
was shown that coherent states on the Lobatchevski space offer minimal entropy production
for the underlying quantum stochastic process \cite{BO}. However, coherent states cannot
be thought of as continuous pointer states. Firstly, although they offer maximum
predictability and are least affected by the environment, they do evolve in time, and,
secondly, they are not orthogonal. Hence, the algebra they generate is neither immune to
the interaction with the reservoir, nor commutative.

We thank R. Alicki for helpful discussions. R.O. acknowledges financial support from
Alexander von Humboldt Foundation.

\end{document}